\documentclass[]{aa}  
\usepackage{graphicx}
\usepackage{txfonts}
\usepackage{subfigure}
\usepackage{natbib}
\def\HII{H\,{\sc ii} }

\begin{document}
   \title{The distribution of warm gas in the G327.3--0.6 massive star-forming region}
   \author{S. Leurini\inst{1}
          \and
          F. Wyrowski\inst{1} \and F.~Herpin\inst{2,3} \and F.~van der Tak\inst{4,5} \and R. G\"usten\inst{1} \and  E.F.~van~Dishoeck\inst{6,7}
\institute{Max-Planck-Institut f\"ur Radioastronomie, Auf dem H\"ugel 69, 53121 Bonn, Germany\\
              \email{sleurini@mpifr.de}
\and Univ. de Bordeaux, LAB, UMR 5804, F-33270 Floirac, France 
\and CNRS, LAB, UMR 5804, F-33270 Floirac, France
\and SRON Netherlands Institute for Space Research, PO Box 800, 9700 AV, Groningen, The Netherlands
\and Kapteyn Astronomical Institute, University of Groningen, PO Box 800, 9700 AV, Groningen, The Netherlands
\and Leiden Observatory, Leiden University, PO Box 9513, 2300 RA Leiden, The Netherlands
\and Max Planck Institut f\"{u}r Extraterrestrische Physik, Giessenbachstrasse 1, 85748 Garching, Germany}}

   \date{\today}

  \abstract
   {}
   {Most studies of high-mass star formation focus on massive and/or luminous clumps, 
but the physical properties of their larger scale environment are poorly known. In this work, we aim at 
characterising the effects of clustered star formation and feedback of massive stars on the surrounding medium by 
studying the distribution of warm gas through mid-$J$ $^{12}$CO and $^{13}$CO observations.}
   {We present APEX $^{12}$CO(6--5), (7--6), $^{13}$CO(6--5), (8--7) and HIFI $^{13}$CO(10--9) maps of the star forming region G327.36--0.6 with a linear size of $\sim 3$~pc~$ \times~4$~pc. 
We infer the physical properties of the emitting gas on large scales through a local thermodynamic equilibrium analysis, while we apply a more sophisticated large velocity gradient approach on selected positions.}
   {Maps of all lines are dominated in intensity by the photon dominated region around the H{\sc ii} region G327.3--0.5. 
Mid-$J$ $^{12}$CO emission is detected over the whole extent of the maps 
with excitation temperatures ranging from 20~K up to 80~K in the gas around the H{\sc ii} region,
and H$_2$ column densities from few $10^{21}$~cm$^{-2}$  in the 
inter-clump gas to $3\times 10^{22}$~cm$^{-2}$ towards the hot core G327.3--0.6.  The warm gas (traced by $^{12}$ and $^{13}$CO(6--5) emission)
is only a small percentage ($\sim$10\%) of the total gas in the infrared dark cloud, while it reaches values up to 
$\sim$35\% of the total gas in the ring surrounding the H{\sc ii} region. The $^{12}$CO ladders
are qualitatively compatible with photon dominated region models for high density gas, but the much weaker than predicted $^{13}$CO emission  
suggests
that it comes from  
 a large number of clumps  
along the line of sight.
All
lines are detected in the inter-clump gas when averaged over a large
region with an equivalent radius of 50\arcsec ($\sim$0.8~pc),  implying that the mid-$J$ $^{12}$CO and $^{13}$CO 
inter-clump emission is due to high
density components with low filling factor. Finally, 
the detection of the $^{13}$CO(10--9) line allows to disentangle the effects of gas temperature 
and gas density on the CO emission, which are degenerate in the APEX observations alone.}
   {}

   \keywords{Stars: formation -- ISM: H{\sc ii} regions -- ISM: individual objects: G327.36--0.6}
   \titlerunning{Warm gas in the G327.3--0.6 massive star-forming region}
   \maketitle
%

\section{Introduction}
The influence of high-mass stars on the interstellar medium is tremendous. During their process of formation,
they are sources of powerful, bipolar outflows \citep[e.g.,][]{2002A&A...383..892B}, their strong ultraviolet and
far-ultraviolet radiation fields give rise to bright H{\sc ii} and photon dominated regions (PDRs) and during their
whole lifetime powerful stellar winds interact with the surroundings.  
Finally, their short life ends in a violent supernova explosion,
injecting heavy elements into the interstellar medium and possibly triggering further star formation with
the accompanying shocks. These are also the type of regions that dominate far-infrared observations of starburst
galaxies.

Most studies of massive star formation focus on emission peaks at infrared or submillimetre wavelengths, 
which correspond to peaks in the temperature and/or mass distribution. The aim of our work is to 
characterise the effects of clustered star formation and feedback of massive stars on the surrounding medium. 
We have made APEX maps of three cluster-forming regions (G327.3--0.6, NGC6334 and W51) in mid-$J$ $^{12}$CO ((6--5) and (7--6)) and $^{13}$CO transitions 
((6--5) and (8--7)) in order  to have a direct measure of the excitation of the warm extended inter-clump gas between 
dense cores in the cluster \citep[see for example][]{1986ApJ...300L..89B,1990ApJ...356..513S}.
Our sample of sources was chosen among six nearby cluster-forming clouds mapped in water 
and in the $^{13}$CO(10--9) transition 
as part of the Water in Star-Forming Regions with Herschel (WISH)  \citep{2011PASP..123..138V} 
guaranteed time key program (GT-KP) for the {\it Herschel} Space Observatory \citep{2010A&A...518L...1P}.

In this paper, we present the $^{12}$CO and $^{13}$CO maps of the star-forming region G327.3--0.6  
 at a distance of 3.3~kpc \citep[][based on H{\sc~I} absorption]{2011MNRAS.tmp.2112U}.
G327.3--0.6  is well suited to
study  cluster-forming clouds because of its relatively close distance and because
several sources in different evolutionary phases coexist in a small region, as found by \citet{2006A&A...454L..91W}. 
 Our maps (with a linear extension of $\sim 3$~pc~$ \times~4$~pc) cover the H{\sc ii} region G327.3--0.5 \citep{1970AuJPA..14....1G} associated with a luminous PDR, and an infrared dark cloud (IRDC)
\citep{2006A&A...454L..91W} hosting the bright hot core G327.3--0.6 \citep{2000ApJ...545..309G} and 
the extended green object (EGO) candidate G327.30--0.58 \citep{2008AJ....136.2391C}. EGOs are identified through their extended 4.5~$\mu$m
 emission in the {\it Spitzer} IRAC2 band, which is believed  to trace outflows  from massive young stellar objects (YSOs) \citep{2008AJ....136.2391C}.  

This paper is organised as follows: in Sect.~\ref{allobs} we present the APEX and {\it Herschel}\footnote{{\it Herschel} is an ESA space observatory with science instruments provided by European-led Principal Investigator consortia and with important participation from NASA.} observations of G327.3--0.6, 
in Sect.~\ref{obs_res} we discuss the morphology and kinematics of the $^{12}$CO and $^{13}$CO emission, in Sect.~\ref{res} we investigate the physical conditions of the emitting gas. Finally, in Sect.~\ref{dis} we discuss our results and compare to similar observations performed towards low- and high-mass star forming regions. Our results are summarised in Sect.~\ref{end}. 

\begin{figure}
\centering
\includegraphics[width=9cm]{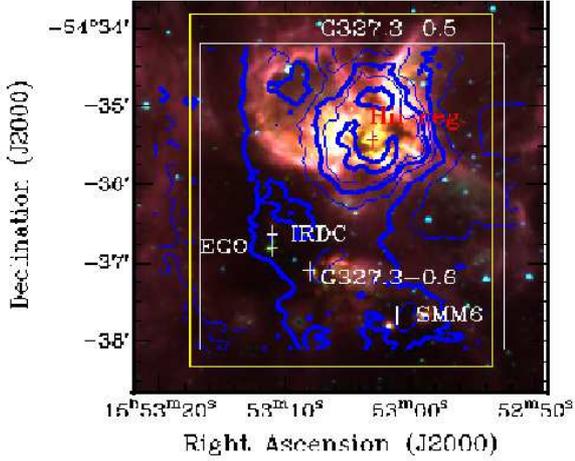}
\caption{{\it Spitzer} infrared colour image of the region G327.3--0.6 with red representing 8.0$\mu$m, 
green 4.5$\mu$m and blue 3.6$\mu$m. The blue contours represent the integrated intensity of the $^{12}$CO(6--5) 
line (thin contours are 15\%, 45\% and 70\% of the peak emission, thick contours 30\%, 60\% and 90\% of the peak emission). The  sources discussed in this paper (the IRDC, the EGO, 
the hot core G327.3--0.6, the SMM6 position  and
the H{\sc ii} region  G327.3--0.5) are also marked with crosses. The white and yellow boxes mark the regions mapped with 
APEX and {\it Herschel}, respectively.}
\label{ir}
\end{figure}

\section{Observations}\label{allobs}
\begin{table}
\centering
\caption{Observational parameters.}\label{obs}
\begin{tabular}{cccccc}
\hline
\hline
\multicolumn{1}{c}{Line} &\multicolumn{1}{c}{Frequency}&\multicolumn{1}{c}{Telescope}&
\multicolumn{1}{c}{Beam}&\multicolumn{1}{c}{$\Delta\varv$}&\multicolumn{1}{c}{r.m.s.}\\
\multicolumn{1}{c}{} &\multicolumn{1}{c}{(GHz)}&&\multicolumn{1}{c}{($\arcsec$)}&
\multicolumn{1}{c}{(km~s$^{-1}$)}&\multicolumn{1}{c}{(K)}\\
\hline
CO(6--5)&691.473&APEX&9\farcs0&1.0&1.4\\
CO(7--6)&806.652&APEX&7\farcs7&1.0&4.4\\
$^{13}$CO(6--5)&661.067&APEX&9\farcs4&1.0&1.1\\
$^{13}$CO(8--7)&881.273&APEX&7\farcs1&1.0&2.8\\
$^{13}$CO(10--9)&1101.350&Herschel&$19\farcs0$&1.1&0.1\\
\hline

\end{tabular}

\end{table}
\subsection{APEX telescope}\label{obs_a}
The CHAMP$^+$ \citep{2006SPIE.6275E..19K,2008SPIE.7020E..25G} dual
colour heterodyne array receiver of 7 pixels per frequency channel on the APEX telescope\footnote{This
  publication is based on data acquired with the Atacama Pathfinder
  Experiment (APEX). APEX is a collaboration between the
  Max-Planck-Institut fur Radioastronomie, the European Southern
  Observatory, and the Onsala Space Observatory} was used in September
2008 to simultaneously map  the star-forming region G327.3--0.6 in the
$^{12}$CO (6--5) and (7--6) lines and, in a second coverage, the
$^{13}$CO (6--5) and (8--7) transitions.  The region from the hot core
in G327.3--006 to the \HII region G327.3--0.5 (Fig.~\ref{ir}) was covered with on-the-fly
maps of 200\arcsec$\times$240\arcsec\ spaced by 4\arcsec\ in
declination and right ascension.  

We used the Fast Fourier Transform Spectrometer \citep[FFTS,][]{2006A&A...454L..29K} as
backend with two units of fixed bandwidth of 1.5 GHz and 8192 channels per pixel.
We used the two IF groups of the FFTS with an offset of $\pm 460$~MHz between them. 
The original resolution of the dataset is $0.3$~km~s$^{-1}$; the spectra were 
smoothed to 1~km~s$^{-1}$ for a better signal-to-noise ratio. 

The observations were performed under good weather conditions
with a precipitable water vapour level between 0.5 and 0.7~mm.
Typical single side band system temperatures during the observations were around 1600~K
and 5200~K, for the low and high frequency channel respectively. The
conversion from antenna temperature units to brightness temperatures
was done assuming a forward efficiency of 0.95 for all channels, and a
main beam efficiency of 0.48 for the $^{12}$CO and $^{13}$CO (6--5)
observations, 0.45 for the $^{12}$CO (7--6) data, and 0.44 for $^{13}$CO(8--7), as measured on Jupiter in September
2008\footnote{http://www.mpifr-bonn.mpg.de/div/submmtech/heterodyne/\\champplus/champmain.html}. 
The pointing was checked on the continuum  of the hot core G327.3--0.6 
($\alpha_{\rm{J2000}}=15^h53^m07^s.8, \delta_{\rm{J2000}}=-54\degr36\arcmin06\farcs4$).
The
maps were produced with the XY\_MAP task of CLASS90\footnote{http://www.iram.fr/IRAMFR/GILDAS/}, which convolves
the data with a Gaussian of one third of the beam: the final angular
resolution is 9$\farcs$4 for the low frequency data, 8$\farcs$1 for
the high frequency.
\subsection{{\it Herschel} Space Observatory}
The $^{13}$CO (10--9) line (see Table~\ref{obs}) was mapped (size=$210\arcsec\times 270\arcsec$) with 
the HIFI instrument \citep{2010A&A...518L...6D} towards G327.3-0.6 on February, 18th, 2011 (observing day (OD) 645, observing identification number 
(OBSID) 1342214421. The centre of the map is 
 $\alpha_{\rm{J2000}}=15^h53^m05^s.48, \delta_{\rm{J2000}}=-54\degr36\arcmin06\farcs2$. The observations 
are part of the WISH GT-KP \citep{2011PASP..123..138V}. Data were taken simultaneously in H and V polarisations using both 
the acousto-optical Wide-Band Spectrometer (WBS) with 1.1 MHz resolution and the correlator-based 
High-Resolution Spectrometer (HRS) with 250 kHz nominal resolution. In this paper we present only the 
WBS data.
We used the on-the-fly mapping
 mode with Nyquist sampling. HIFI receivers are double sideband with a sideband ratio close to unity. 
The double side band system temperatures and total integration times are respectively
 384~K and 3482~s. The rms noise level at 1~km~s$^{-1}$ spectral resolution is $\sim$0.1~K. Calibration of the raw data onto $T_A$ scale was performed by the 
in-orbit system \citep{2012A&A...537A..17R}; conversion to $T_{mb}$ was done with a beam efficiency of 0.74 and a 
forward efficiency of  0.96. The flux scale accuracy is estimated to be around 15\% for band 3. 
Data calibration was performed in the {\it Herschel} Interactive Processing Environment \citep[HIPE,][]{2010ASPC..434..139O} version 6.0. 
Further analysis was done within the CLASS90 package. After inspection, 
data from the two polarisations were averaged together.

The original angular resolution of the data is 19\farcs0. The final
maps were produced with the XY\_MAP task of CLASS90 and have  an angular
resolution of 21\farcs1.
\section{Observational results}\label{obs_res}

\subsection{Morphology}
\begin{figure*}
\centering
\includegraphics[angle=-90,width=18cm]{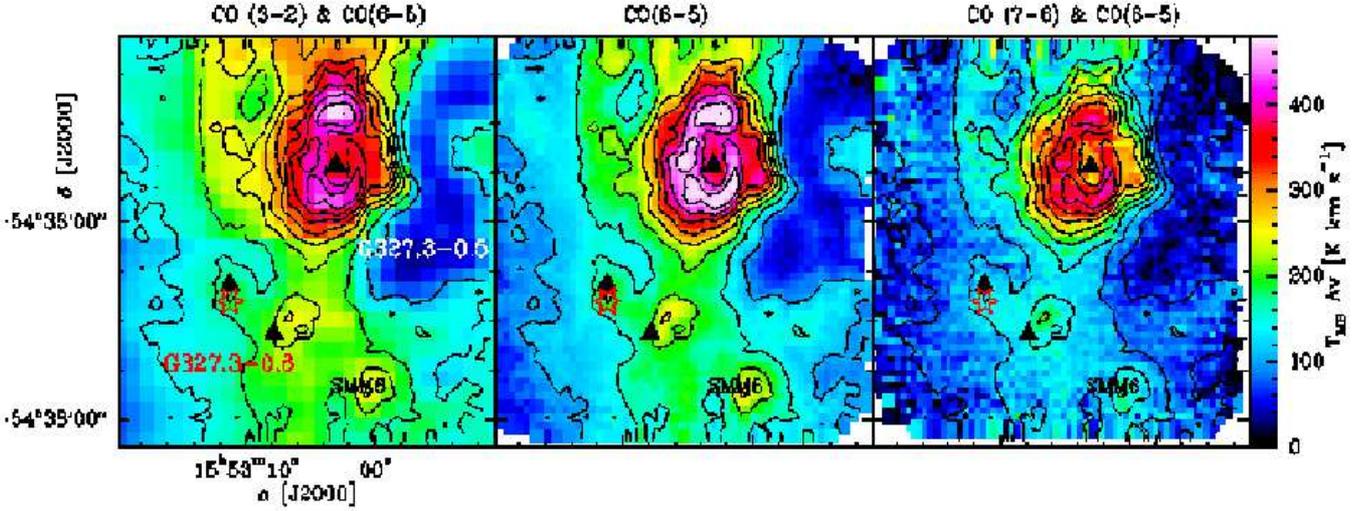}
\caption{Maps of the integrated intensity of the $^{12}$CO(3--2), (6--5) and (7--6) lines in the velocity range 
$\rm{v_{\rm{LSR}}}=[-54,-40]$~km~s$^{-1}$ (colour scale). Solid contours are the integrated intensity of the $^{12}$CO(6--5) line from 20\% of the peak emission in steps of 10\%.
In each panel, the positions analysed in Sect.~\ref{sed} (the hot core, the IRDC position, and
the centre of the H{\sc ii} region)
are marked with black triangles. The red star labels the position of the EGO. SMM6 (left panel) is one of the submillimetre sources detected by \citet{2009A&A...501L...1M}.}\label{co}
\end{figure*} 
\begin{figure*}
\centering
\includegraphics[angle=-90,width=18cm]{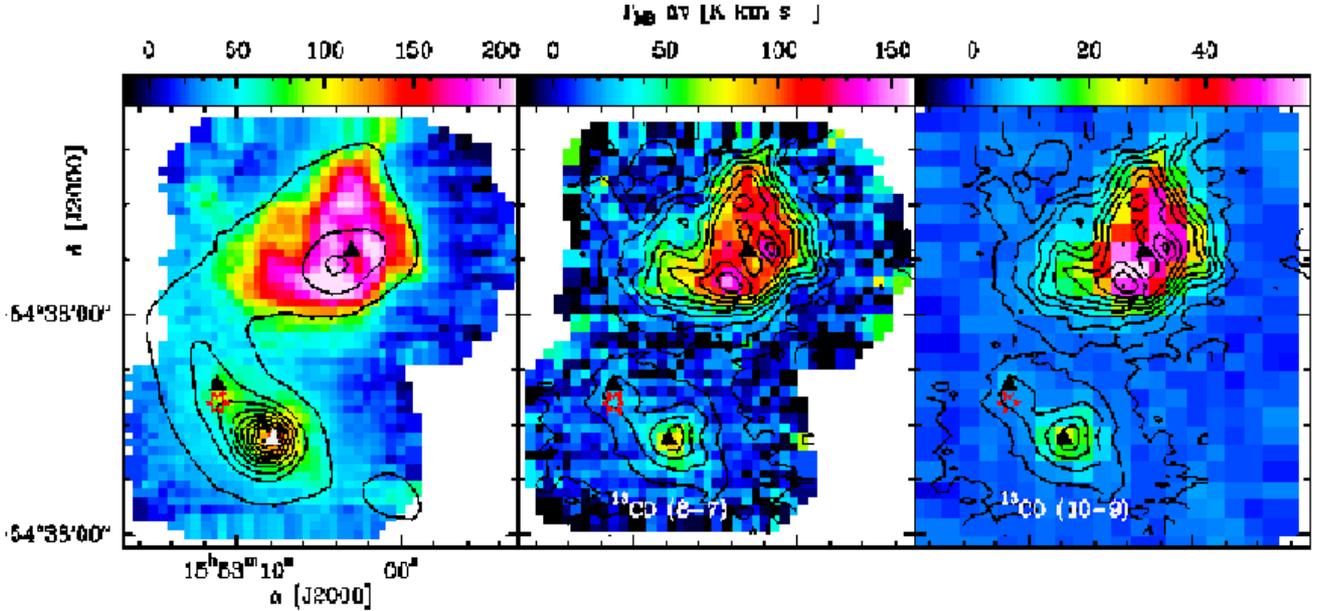}
\caption{Maps of the integrated intensity of the $^{13}$CO(6--5), (8--7) and (10--9) transitions 
in the velocity range $\rm{v_{\rm{LSR}}}=[-54,-40]$~km~s$^{-1}$ (colour scale). 
The solid contours in the left panel represent the LABOCA continuum emission at 870$\mu$m from 5\% of the peak emission in steps of 10\% \citep{2009A&A...504..415S}. In the middle and right panels, the solid contours are 
the integrated intensity of the $^{13}$CO(6--5) from 20\% of the peak emission in steps of 10\%.
In each panel, the positions analysed in Sect.~\ref{sed} (the hot core, the IRDC position, and
the centre of the H{\sc ii} region)
are marked with black triangles (except in the left panel, where the hot core is shown by a white triangle). The red star labels the position of the EGO.}\label{13co}
\end{figure*} 

Figure~\ref{ir} shows  the $^{12}$CO(6--5) integrated intensity map overlaid on the composite image of the  
IRAC {\it Spitzer} 3.6,  4.5 and 8.0$\mu$m bands of the region. The $^{12}$CO emission traces the distribution 
of the 8.0$\mu$m emission, but it is also associated with the infrared dark cloud found on the east of the hot core.
The EGO candidate  G327.30--0.58 identified by \citet{2008AJ....136.2391C} and clearly visible in Fig.~\ref{ir},
is also detected in the $^{12}$CO data as secondary peak of emission (Fig.~\ref{co}).
The map of the 
integrated intensities of the $^{12}$CO(7--6) line is also presented in Fig.~\ref{co} together 
with the integrated intensity of the  $^{12}$CO(3--2) line from \citet{2006A&A...454L..91W}. Figure~\ref{13co} shows the distribution of the 
 $^{13}$CO(6--5), (8--7) and (10--9) emissions. 
The accuracy of the
relative pointings was checked on the hot core G327.3--0.6. For this purpose, we derived integrated intensity maps
of lines detected only towards this position, which are close in frequency to the 
 $^{12}$CO(6--5), $^{13}$CO(6--5) and  $^{13}$CO(8--7) transitions, and therefore were observed simultaneously to
the current dataset. From these data, we infer
a position for the hot core in agreement
with interferometric measurements at 3~mm \citep[][and Table~\ref{pos}]{2008Ap&SS.313...69W}
within $\sim$1\farcs5.

\begin{table}
\centering
\caption{Coordinates of the main sources in the G327.3--0.6 massive star-forming region}\label{pos}
\begin{tabular}{lcc}
\hline
\hline
\multicolumn{1}{c}{source} &\multicolumn{1}{c}{$\alpha_{\rm{J2000}}$}&\multicolumn{1}{c}{$\delta_{\rm{J2000}}$}\\
\hline
SMM6$^{\rm{a}}$&15:53:00.9&-54:37:40.0\\
hot core$^{\rm{b}}$&15:53:07.8&-54:37:06.4\\
EGO$^{\rm{c}}$&15:53:11.2&-54:36:48.0\\
H{\sc ii}$^{\rm{d}}$&15:53:03.0&-54:35:25.6\\

\hline

\end{tabular}
\begin{list}{}{}
\item[$^{\rm{a}}$] \citet{2009A&A...501L...1M}
\item[$^{\rm{b}}$] \citet{2008Ap&SS.313...69W}
\item[$^{\rm{b}}$] \citet{2008Ap&SS.313...69W}
\item[$^{\rm{c}}$] \citet{2008AJ....136.2391C}
\item[$^{\rm{d}}$] peak of the centimetre continuum emission from ATCA archival data at 2.3~GHz, project number C772
\end{list}
\end{table}

All observed $^{12}$CO transitions trace the H{\sc
  ii} region G327.3--0.5 as well as the infrared dark cloud 
which hosts the hot core G327.3--0.6. Moreover, the $^{12}$CO(6--5), (7--6) and $^{13}$CO(6--5) lines show 
extended emission along a ridge running approximately N-S that matches very well with the distribution of the
CO(3--2) transition.
The hourglass shape hole to the west of the
H{\sc ii} region G327.3--0.5 where the $^{12}$CO(3--2) emission is strongly
reduced \citep[see][]{2006A&A...454L..91W} is seen also in the $^{12}$CO(6--5) and (7--6) lines which, although
much weaker than in the rest of the map, are still detected at this
position.

All transitions peak towards the H{\sc ii} region G327.3--0.5 where the
main isotopologue lines have intensities up to 60--65~K. The 
integrated intensities of the CO isotopologues show a distribution along a
ring-like structure around the  peak of the cm continuum emission
\citep{1970AuJPA..14....1G}. The centre of the ring also coincides
with the massive young stellar object number 87 identified in the near-infrared by \citet{2011MNRAS.411..705M}.
Since the ring is detected also in high-$J$
transitions of $^{13}$CO, it is plausible that this morphology is true and
not due to optical depth effects. This structure likely coincides with the limb brightening of the
hot surface of a PDR around G327.3--0.5 and
could trace an expanding shell. We will investigate this scenario in Sect.~\ref{velo}.

The hot core G327.3--0.6 shows up as a secondary peak in the integrated
intensity maps of the $^{13}$CO transitions, while the main CO
isotopologue peaks to its north-west, probably because of  optical depth effects. Strong
self-absorption profiles 
 are indeed detected  in all $^{12}$CO
 lines towards the hot core (see Sect.~\ref{velo}). The submillimetre source SMM6 \citep[seen in the continuum emission at 450$\mu$m by][]{2009A&A...501L...1M} is detected as a peak of emission in all 
integrated intensity maps of $^{12}$CO and in $^{13}$CO(6--5), although at the edge of the 
mapped region. The other submillimetre sources are also marked in Figs.~\ref{co}. 
The EGO candidate  G327.30--0.58
is also detected in the $^{13}$CO(6--5) map (Fig.~\ref{13co}). The $^{13}$CO(6--5) traces the whole IRDC and
not only the active site of star formation where the EGO is detected.
The continuum emission due to dust (seen for example at 870$\mu$m in Fig.~\ref{13co}) follows the  distribution
of the $^{13}$CO lines.

\begin{figure}
\centering
\includegraphics[angle=-90,width=9cm]{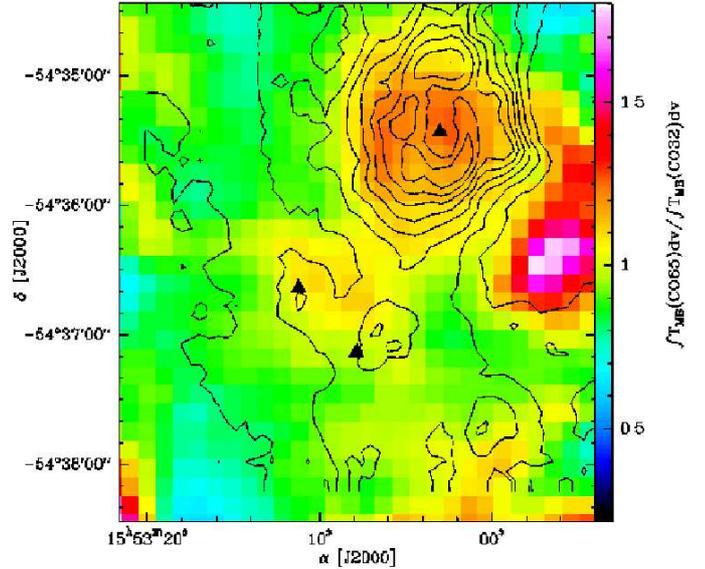}
\caption{Distribution of the line ratio of the $^{12}$CO(6--5) transition to the $^{12}$CO(3--2) line. 
Solid contours show the $^{12}$CO(6--5) integrated intensity in the velocity range 
$\rm{v_{\rm{LSR}}}=[-54,-40]$~km~s$^{-1}$ from 20\% of the peak value in steps of 10\%. The black triangles are as in 
Fig.~\ref{co}.}
\label{ratioco32}
\end{figure}

In Fig.~\ref{ratioco32} we show the ratio 
of the integrated intensity of the $^{12}$CO(6--5) transition (convolved to the 18\arcsec~resolution of the $^{12}$CO(3--2) data) to that
of the $^{12}$CO(3--2) line. This ratio ranges between 0.3 and 1.8; it has values slightly larger than   one 
towards the H{\sc ii} region (1.2 at its centre),  while it is about unity towards the hot core. The peak is found south-west of the H{\sc ii} region G327.3--0.5, where both lines are detected with a high confidence level.
However, these results could be biased by the strong self absorption in both $^{12}$CO lines 
(see Sect.~\ref{velo}). For this reason, we 
computed the ratio between the two transitions  in four 
 velocity ranges to cross-check the results of Fig.~\ref{ratioco32}. The inferred values, however, do not change significantly. 

\subsection{Line profiles and velocity field}\label{velo}
\begin{figure}
\centering
\includegraphics[angle=-90,width=9cm]{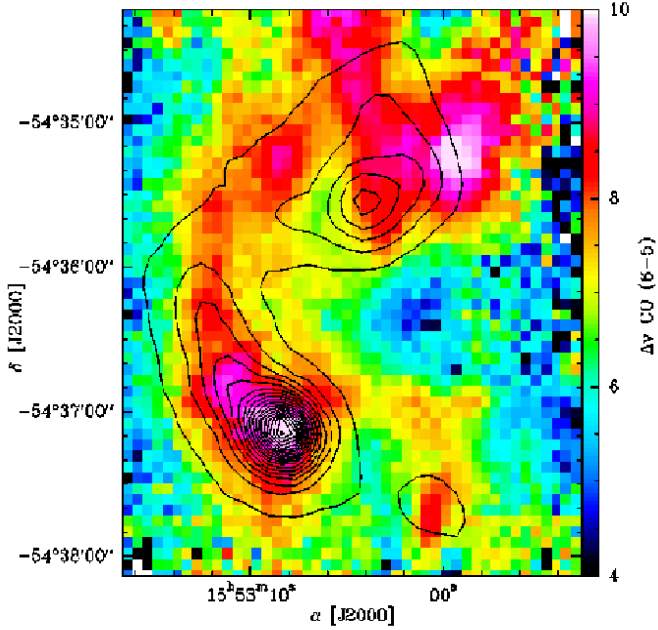}
\caption{Distribution of the second  moment of the $^{12}$CO(6--5) line. The black solid contours represent the 
LABOCA continuum emission at 870$\mu$m from 5\% of the peak emission in steps of 5\%. The peak of the continuum 
emission corresponds to the hot core position.}
\label{width}
\end{figure}

The widespread $^{12}$CO(6--5) emission shows line profiles with a typical
width of $\sim$8~km~s$^{-1}$ in the gas between G327.3--0.5 and the
infrared dark cloud. Broader profiles are detected in the infrared
dark cloud and in the northern part of the H{\sc ii} region
G327.3--0.5. Figure~\ref{width} shows the distribution of the line width of the $^{12}$CO(6--5) transition: 
the $^{12}$CO(6--5) line width follows an arc-like 
structure that connects the H{\sc ii} region G327.3--0.5 to the infrared dark cloud where the hot core is. 
Interestingly, the same morphology is seen in the LABOCA map of the region \citep{2009A&A...504..415S}. Line widths are similar for all $^{12}$CO lines, while
they are consistently narrower in the $^{13}$CO transitions. 

\begin{figure}
\centering
\includegraphics[angle=-90,width=7cm]{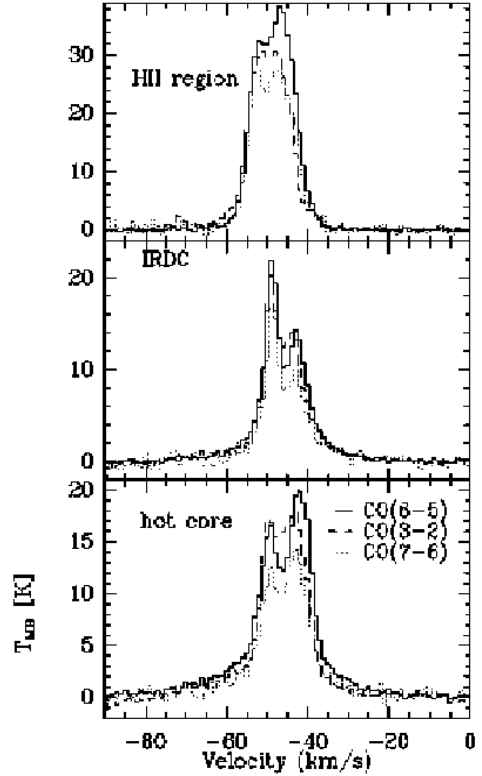}
\caption{Mean spectra over a beam of 18$''$ (to match the resolution of the $^{12}$CO(3--2) data) of the $^{12}$CO isotopologue transitions analysed in the paper. The top panel shows
spectra at the centre of G327.3--0.5, the middle panel spectra at the IRDC position, the bottom panel spectra at the hot core position.}
\label{spectra}
\end{figure}

\begin{figure}
\centering
\includegraphics[angle=-90,width=7cm]{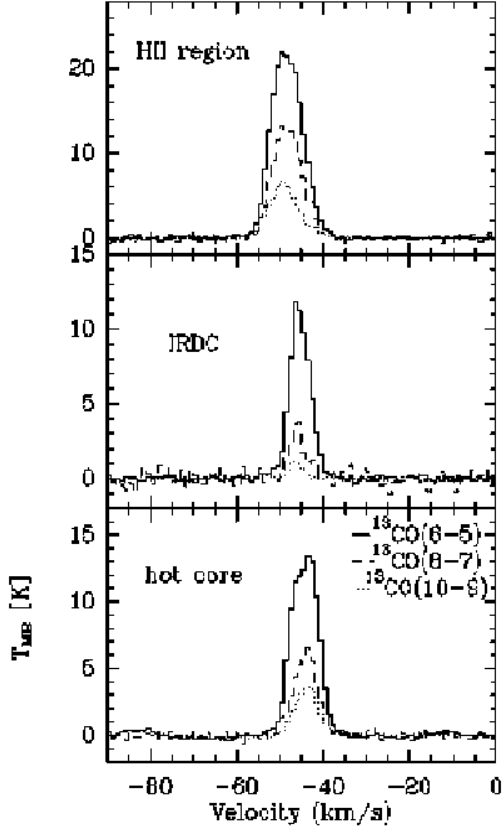}
\caption{Mean spectra over a beam of 21\farcs1 (to match the resolution of the $^{13}$CO(10--9) data) 
of the $^{13}$CO isotopologue transitions. 
The selected positions are those discussed in 
Sect.~\ref{sed}.}
\label{spectra_13co}
\end{figure}

\begin{figure}
\centering
\includegraphics[angle=-90,width=8cm]{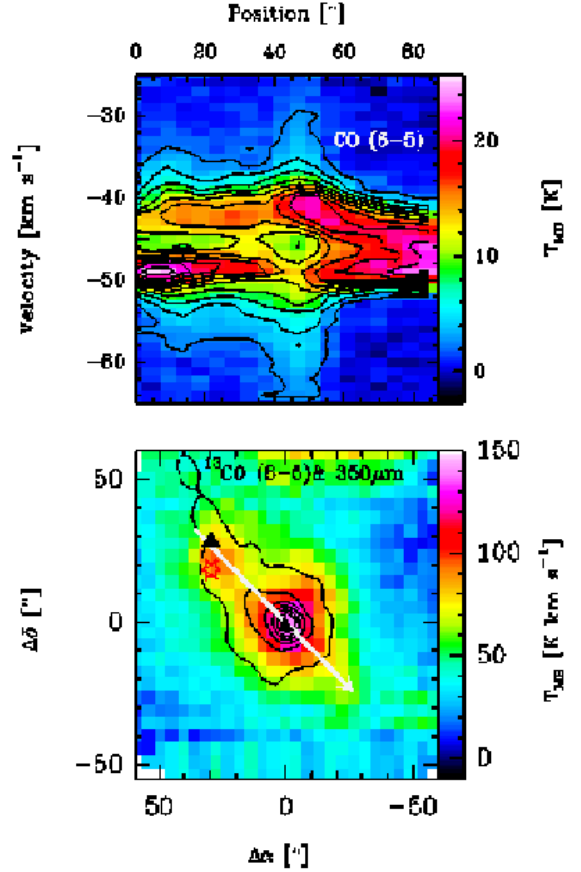}
\caption{{\bf Top panel:} Colour scale and contours show the P-V diagram of the 
CO(6--5) transition computed along the cut indicated by the white arrow in the bottom panel. Offset positions increase along the direction of the arrow shown in the bottom panel. {\bf Bottom panel:} distribution of the integrated intensity of the $^{13}$CO(6--5) transition 
towards the hot core G327.3--0.6. Solid contours show the 
continuum emission at 350$\mu$m (Wyrowski et al., in prep.) from 3$\sigma$ in steps of 10$\sigma$ ($\sigma\sim 3$~Jy/beam).  
 Symbols are as in Fig.~\ref{co}.}
\label{hc}
\end{figure}

\begin{figure}
\centering
\includegraphics[angle=-90,width=8cm]{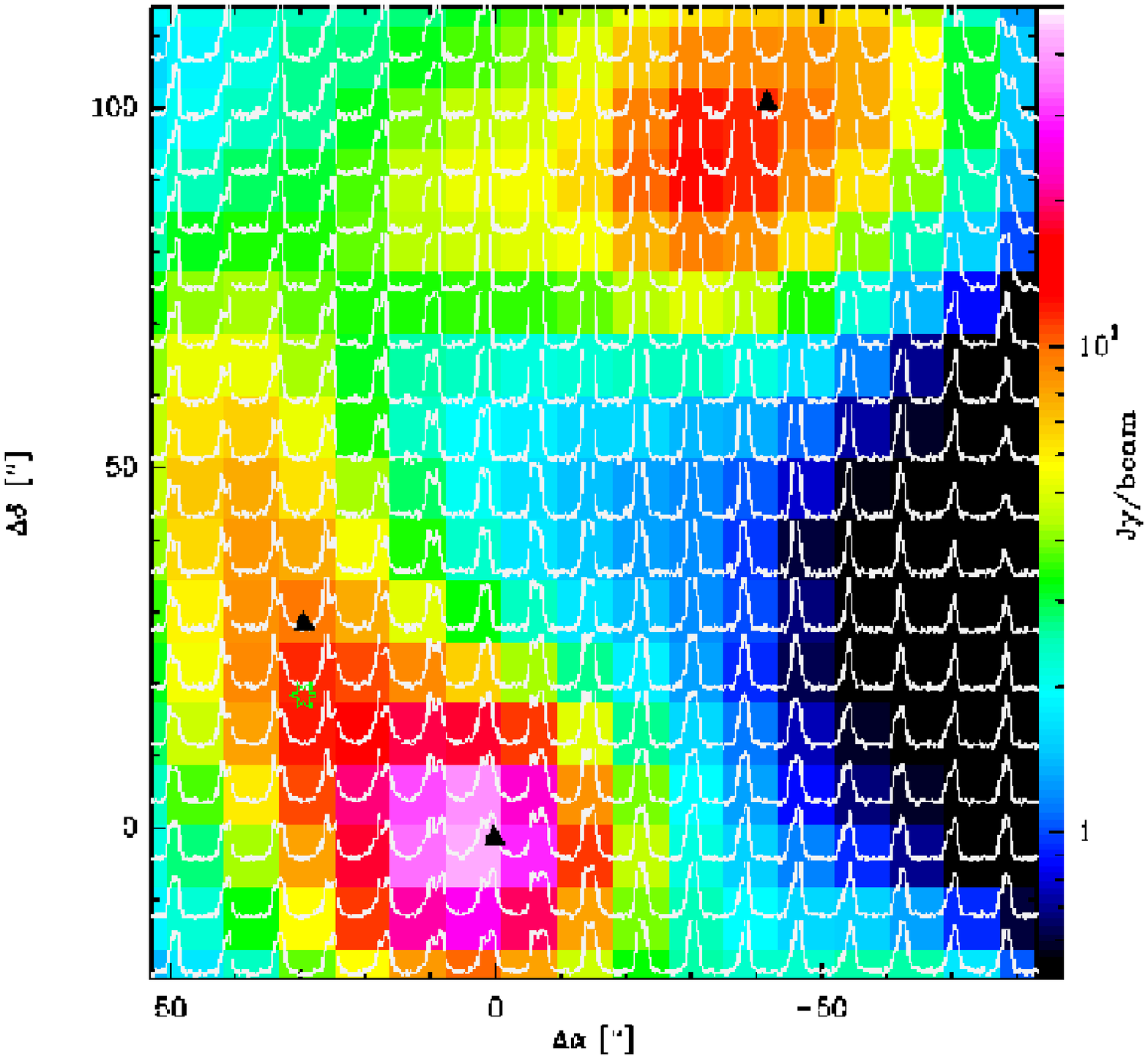}
\caption{Map of the $^{12}$CO(7--6) line overlaid on the continuum emission at 870$\mu$m from LABOCA. The velocity axis ranges
from -60 to -25~km~s$^{-1}$, the temperature axis from -1 to 15~K. The $^{12}$CO(7--6) data were smoothed to a resolution of 18$\arcsec$ 
to match the resolution of the $^{12}$CO(3--2) data and of the LABOCA emission. The centre of the map is that of the APEX data (see Sect.~\ref{obs_a}). The triangles and the green star are as in Fig.~\ref{co}.}\label{co76}

\end{figure}

Representative spectra of all $^{12}$CO transitions analysed in this study
are presented in Fig.~\ref{spectra} towards the hot core, the IRDC position ((30\arcsec, 30\arcsec) 
from the centre of the APEX maps, see Sect.~\ref{sed} and Figs.~\ref{co}-\ref{13co}) and the peak
of the cm continuum emission in G327.3--0.5. Spectra of the $^{13}$CO transitions are shown in Fig.~\ref{spectra_13co}.
 Red- and blue-shifted
 wings are detected in the $^{12}$CO lines in a velocity range between -71 and -24~km~s$^{-1}$ (in $^{12}$CO(6--5)) 
towards 
G327.3--0.6 probably due to outflow motions.  However, 
no sign of bipolar outflows is found when inspecting the integrated intensity maps of 
the blue- and red-shifted wings nor in position-velocity diagrams (Fig.~\ref{hc}). Moreover, very
similar broad lines are detected along the whole extent
of the infrared dark cloud, as shown in the top panel of Fig.~\ref{hc}. At the IRDC position, the wings in $^{12}$CO(6--5)
range from -60 to -31~km~s$^{-1}$.
All main isotopologue transitions analysed in this paper are affected
by self-absorption (see Fig.~\ref{spectra} for reference spectra towards
the hot core, the IRDC position and the H{\sc ii} region); moreover,
even the $^{13}$CO(6--5) line shows weak evidence of self-absorbed profile towards
the hot core. Figure~\ref{co76} shows the $^{12}$CO(7--6) spectra overlaid on the continuum 
emission at 870$\mu$m: the self-reversed profile is spread over a large
area and seems to follow the thermal dust continuum emission.


\begin{figure}
\centering
\includegraphics[width=10cm,angle=-90]{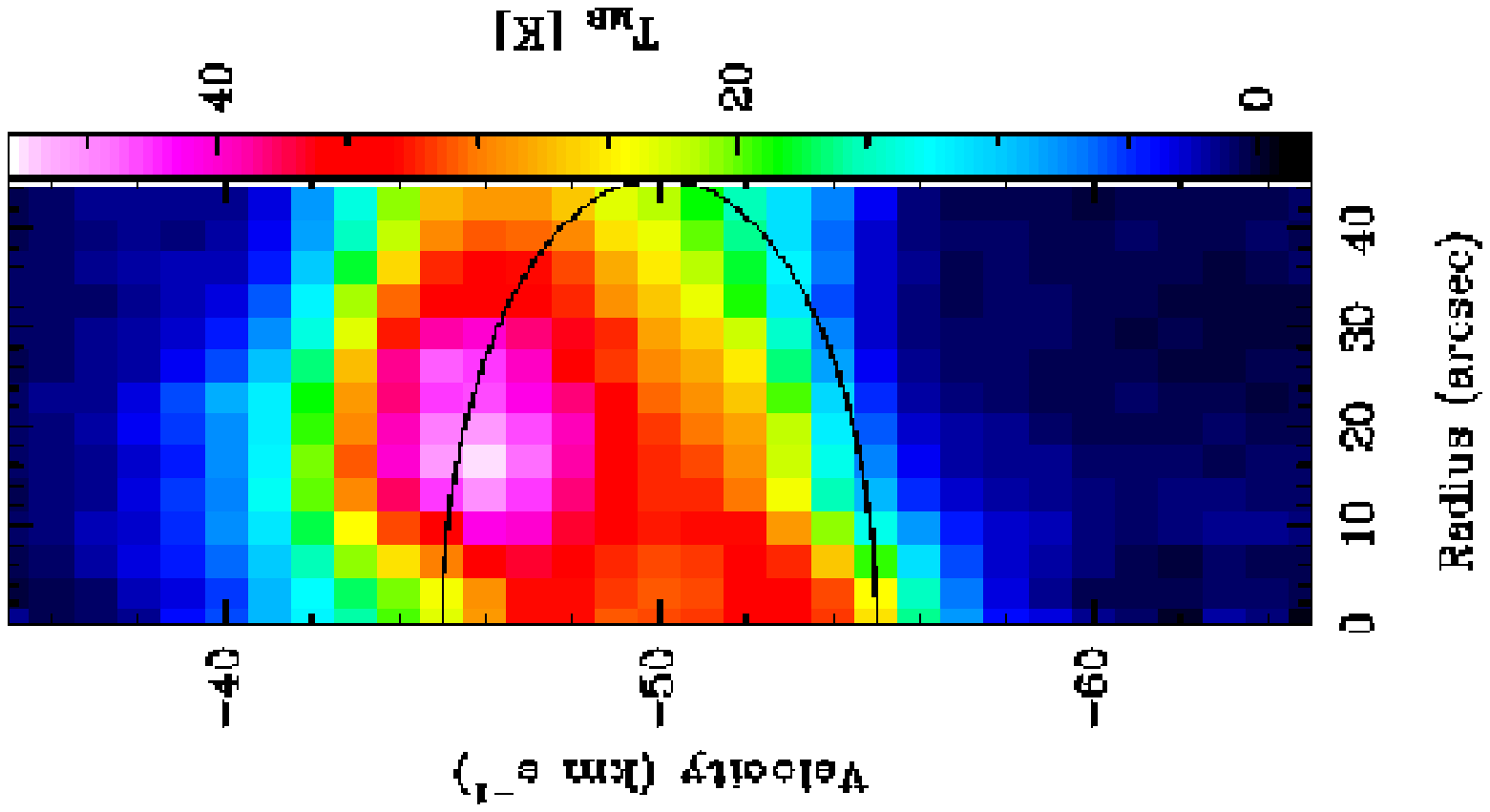}
\caption{{\it (r-\rm{v})} diagram of the H{\sc ii} region G327.3--0.5 obtained from the $^{12}$CO(6--5) data cube. The radius axis is the distance 
to the shell expansion centre, chosen to be the peak of the cm continuum emission.  
The half ellipse represents an ideal shell in  {\it (r-\rm{v})} diagram with an expansion velocity of 5~km~s$^{-1}$.}
\label{shell}
\end{figure}

Finally, the velocity field of the $^{12}$CO transitions may help us to understand
the nature of the ring detected
towards the H{\sc ii} region G327.3--0.5. 
We therefore used the task KSHELL built
in the visualisation software package KARMA
\citep{1996ASPC..101...80G}.  KSHELL computes an average brightness
temperature on annuli about a user defined centre.  
A spherically symmetric expanding shell will look like a half ellipse in a
 {\it (r-\rm{v})} diagram with the axis in the $v$ direction twice the
expansion velocity.  
Figure~\ref{shell} shows the resulting {\it (r-\rm{v})} 
diagram obtained with the $^{12}$CO(6--5) data cube using the peak of 
the cm continuum emission as centre. 
The emission does not follow a perfect spherical shell. This is likely due to 
inhomogeneities in the distribution of the gas, as already seen 
in Fig.~\ref{tex} where the distribution of the optical depth of $^{13}$CO is not symmetric.

\section{Physical conditions of the warm gas}\label{res}
\subsection{LTE analysis}\label{lte}
From the line ratio of the $^{12}$CO(6--5)  to  $^{13}$CO(6--5) transitions we can
derive the optical depth of the $^{12}$CO(6--5) line emission, which can be
then used to infer the excitation temperature of the line and the
column density of $^{12}$CO in the region.

The line intensity  in a given velocity channel of a given transition  is

\begin{equation}\label{tl}
T_{\rm{L}}=\eta\times[F_\nu(T_{\rm{ex}})-F_\nu(T_{\rm{cbg}})]\times\left(1-e^{-\tau_\nu}\right)
\end{equation}
where $\eta$ is the beam filling factor (assumed to be 1 in the following analysis), 
$F_\nu=h\nu/k\times[\rm{exp}(h\nu/kT)-1]^{-1}$, $T_{\rm{cbg}}=2.7$~K, and $\tau_\nu$ is the optical depth. Under the
local thermodynamic equilibrium (LTE) assumption,  $T_{ex}$ is assumed to be equal to the kinetic temperature of the gas and equal for all transitions. 
In the
following analysis, we study the peak intensities of the $^{12}$CO(6--5) and $^{13}$CO(6--5) lines, and include only the cosmic background  as
background radiation and neglect, for example, any contribution from
infrared dust emission since we do not have any map of the
distribution of the dust temperature. This most likely affects  only
our estimates at the hot core position and possibly towards the H{\sc ii} 
region G327.3--0.5 where SABOCA continuum emission at 350$\mu$m
is also detected (Wyrowski et al., in prep.). For an appropriate analysis of the emission from the hot core, 
see \citet{2011A&A...527A..68R}.

Assuming that the $^{12}$CO(6--5) emission is optically thick and that the $^{12}$CO(6--5) and 
$^{13}$CO(6--5) lines have the same excitation temperatures, 
the optical depth of the $^{13}$CO(6--5) transition, $\tau_{^{13}\rm{CO}}$, is 
\begin{equation}
\tau_{^{13}\rm{CO}}=-ln\left(1-\frac{T_L(^{13}\rm{CO})}{T_L(^{12}\rm{CO})}\right)
\end{equation} 

\begin{figure}
\centering
\includegraphics[angle=-90,width=8cm]{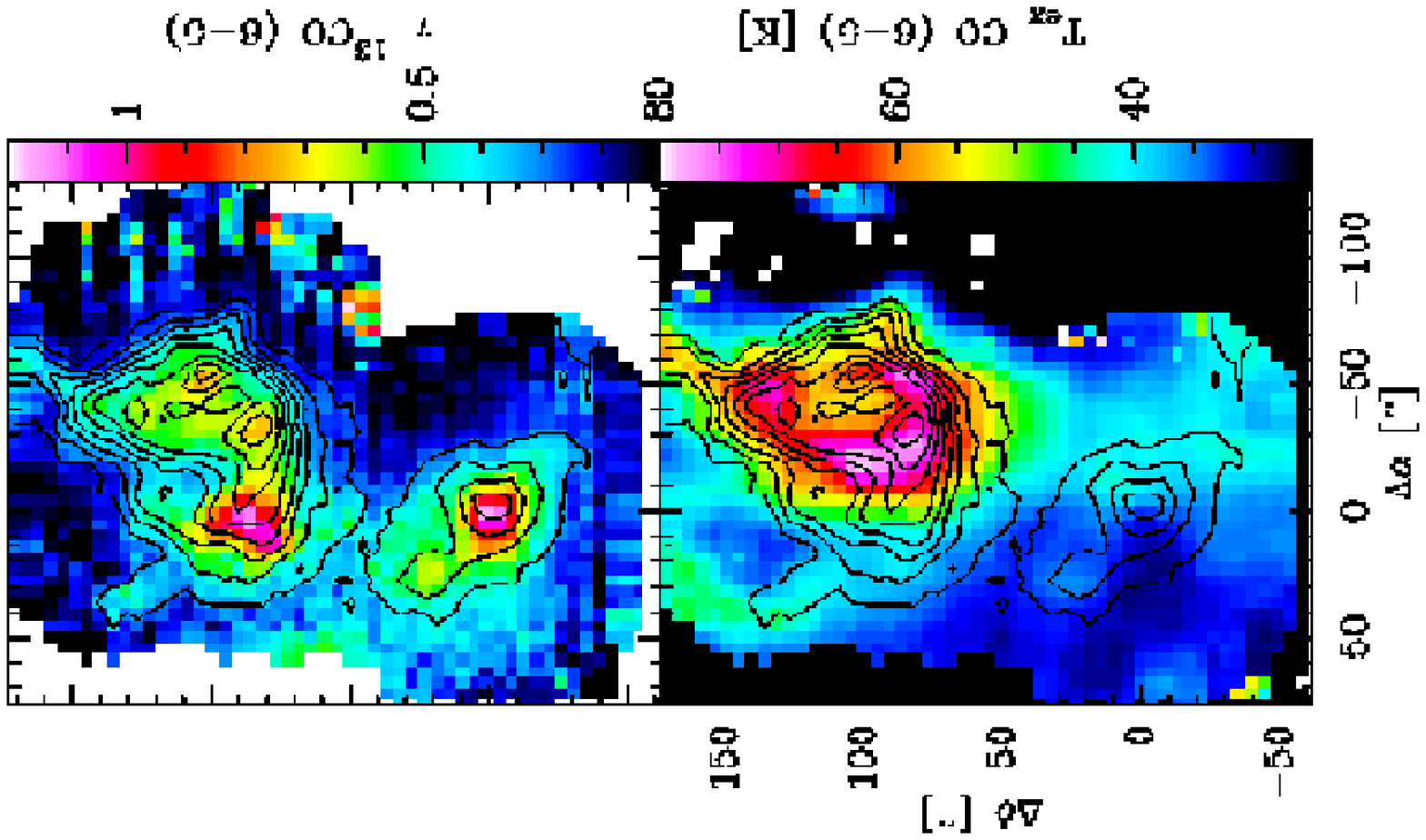}
\caption{Distribution of the optical depth of the $^{13}$CO(6--5) line ($\tau_{^{13}\rm{CO}}$, top panel) and
of the excitation temperature of $^{12}$CO(6--5) transition ($T_{\rm{ex}\rm{CO}}$, bottom panel). Black contours are the $^{13}$CO(6--5) integrated intensity as in Fig.~\ref{13co}.}
\label{tex}
\end{figure}

The optical depth of the $^{12}$CO(6--5) transition can  then be obtained by multiplying 
for the abundance of $^{12}$CO relative to $^{13}$CO, $X_{\rm{^{12}CO}/\rm{^{13}CO}}\sim 60$ \citep{1994ARA&A..32..191W}. From the optical depth of the $^{12}$CO(6--5) line, 
one can also derive its excitation temperature using Eq.~\ref{tl}.
Figure~\ref{tex} shows the distribution of the optical depth of the  
$^{13}$CO(6--5) line and of the excitation temperature of $^{12}$CO(6--5). 
The $^{13}$CO(6--5) emission is moderately optically thick (0.6--0.7) 
at the H{\sc ii} region and at the infrared dark cloud, while it reaches 
values of $\sim$1.2 at the hot core position and in a small part of ring around the H{\sc ii} region.
The map distribution of the excitation temperature of the $^{12}$CO(6--5) line is shown in the bottom
panel of Fig.~\ref{tex}. The map is dominated by the H{\sc ii} region, where $T_{\rm ex}$ reaches values of
 80~K in the ring around the  H{\sc ii} region  and then decreases with increasing distance from it.
The hot core and the rest of the infrared dark cloud have values around 30-35~K.
The excitation temperature increases to the south west of the hot core, in a region where there is also 8$\mu$m emission,
and to the north-east of the H{\sc ii} along a layer of gas also visible in the $^{12}$CO(6--5) integrated intensity map 
(see Fig.~\ref{co}), but more  prominent in the $T_{\rm ex}$ map and in the 8$\mu$m emission map (see Fig.~\ref{ir}).

\begin{figure}
\centering
\includegraphics[angle=-90,width=9cm]{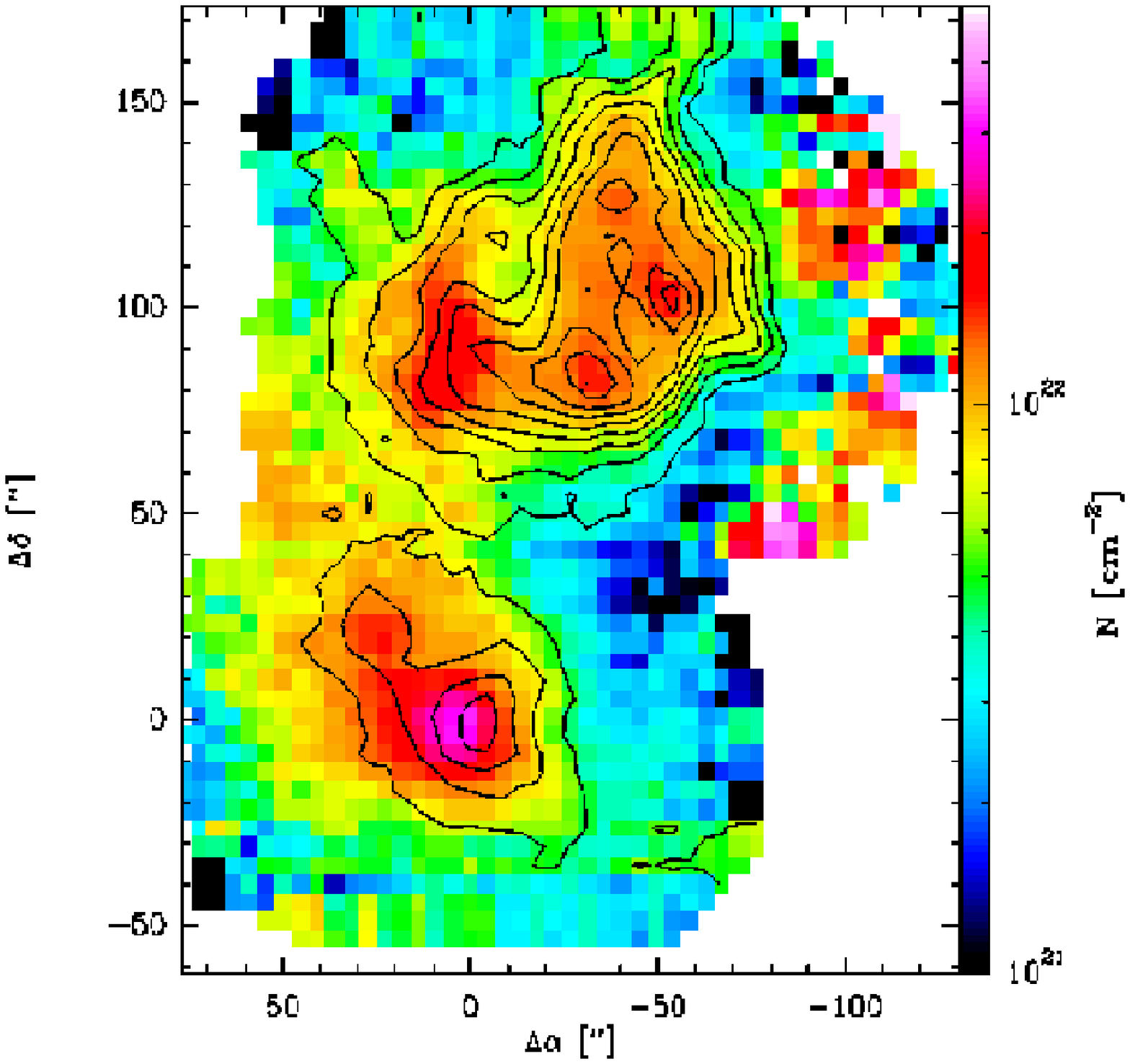}
\caption{Distribution of the H$_2$ column density in the G327.3--0.6 star-forming region based on equation~\ref{tl}. 
Black contours are the $^{13}$CO(6--5) integrated intensity as in Fig.~\ref{13co}.}
\label{colden}
\end{figure}

From the optical depth and the excitation temperature of the $^{12}$CO(6--5) line, we derived the H$_2$ column density 
assuming a relative abundance of $^{12}$CO relative to H$_2$ of $2.7\times10^{-4}$ \citep{1994ApJ...428L..69L}. Results are shown in Fig.~\ref{colden}. 
The largest column density is found towards the hot core ($\sim$$3\times10^{22}$~cm$^{-2}$ in the  9\farcs4 beam of the $^{13}$CO(6--5) data) and decreases 
along the infrared dark cloud with a distribution  similar to that of
the 870$\mu$m continuum emission. 
Three peaks around $10^{22}$~cm$^{-2}$ are found in  the H{\sc ii} region. 

We cross-checked our results by computing the 
H$_{2}$ column density under the assumption that the $^{13}$CO emission is optically thin. The results are consistent with those
presented in Fig.~\ref{colden};  the largest differences (of the order of 30\%) are found towards those positions where the
optical depth of the $^{13}$CO(6--5) transition (Fig.~\ref{tex}) exceeds $\sim$0.7. The assumption of optically thin emission
for $^{13}$CO may be particularly useful for the inter-clump medium (arbitrarily defined as the region in the map where the $^{13}$CO lines are not 
detected on individual spectra), towards which we infer H$_2$ column densities of 
$\sim$$2\times 10^{21}$~cm$^{-2}$ corresponding to a $^{13}$CO column density of $\sim$$10^{16}$~cm$^{-2}$ 
(see also Sect.~\ref{tot}).

We also computed column densities and rotational temperatures in the region using the rotational diagram technique applied
to the $^{13}$CO data. We did not include the $^{12}$CO lines in the analysis because of their complex line profiles and
high optical depths. Given the optical depth previously derived for the $^{13}$CO(6--5) line, we did not 
apply any correction due to optical depth effects to the $^{13}$CO data. 
Results are consistent with the estimates based on equation~\ref{tl}. 
The main differences between the two analyses are found 
towards the hot core, where the rotational temperature is higher than the  excitation derived with 
equation~\ref{tl} ($T_{\rm{rot}}\sim$70~K and $T_{\rm{ex}}\sim$32~K)
and the H$_2$ column density lower ($10^{22}$~cm$^{-2}$ versus $3\times 10^{22}$~cm$^{-2}$ obtained with the first method).
The differences between the two methods are likely
influenced by the self-absorption profile detected in the $^{12}$CO(6--5) line which
results in lower line intensities towards the regions of 
large column densities. In particular, while the first approach 
overestimates
the optical depths (and hence column density) because
of the self absorption, the column density derived with
the rotational diagram analysis is likely underestimated 
towards the hot core because of the optically thin emission assumption, and 
should be corrected by a factor 
$\tau_{\rm{LTE}}/(1-\rm{exp(\tau_{\rm{LTE}})-})\sim$1.7.

\begin{figure}
\centering
\includegraphics[angle=-90,width=9cm]{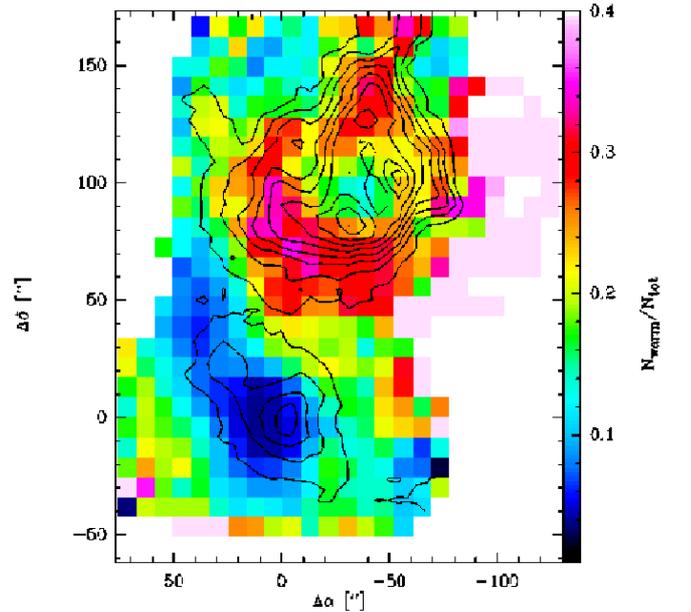}
\caption{Distribution of the ratio between the column density of warm gas (traced by $^{12}$CO and $^{13}$CO(6--5)) and the total 
H$_2$ column density (traced by the continuum emission at 870$\mu$m) in the G327.3--0.6 star-forming region. 
Black contours are the $^{13}$CO(6--5) integrated intensity as in Fig.~\ref{13co}.}
\label{ratio}
\end{figure}

Finally, we  
computed the ratio between the amount of
 warm gas (traced by $^{12}$CO and $^{13}$CO(6--5) and shown in Fig.~\ref{colden}) and 
the total amount of gas (traced by the continuum emission at 870$\mu$m) 
as derived assuming a dust temperature equal to the excitation temperature of $^{12}$CO(6--5) and a dust opacity of 0.0182 cm$^2$~g$^{−1}$ \citep{2008A&A...487..993K}.  
The results are shown in Fig.~\ref{ratio}: the warm gas 
is only a small percentage ($\sim$10\%) of the total gas in the infrared dark cloud, while it reaches values up to 
$\sim$35\% of the total gas in the ring surrounding the H{\sc ii} region.

\subsection{$^{13}$CO ladder}\label{sed}

Since the G327.3--0.6 region was mapped in three different transitions of the $^{13}$CO molecule, we can perform
a multi-line analysis towards selected positions and infer the parameter of the gas. The advantage of  $^{13}$CO 
 compared to the main isotopologue is in the lower opacities of the lines and in the less complex line profiles. 
For this analysis, we used the RADEX program \citep{2007A&A...468..627V} with expanding sphere geometry. 
The molecular dataset comes from the LAMDA database \citep{2005A&A...432..369S} and includes collisional rates adapted
from \citet{2010ApJ...718.1062Y}. We ran models with temperatures from 20 to 200 K, densities in the range 10$^4-5\times10^7$~
cm$^{-3}$, and $^{13}$CO column densities between 10$^{14}$ and 10$^{19}$~cm$^{-2}$.

\begin{table}
\centering
\caption{Line parameters of the $^{13}$CO lines}\label{para}
\begin{tabular}{cccc}
\hline
\hline
\multicolumn{1}{c}{Position}&\multicolumn{1}{c}{v$_{\rm{LSR}}$}&\multicolumn{1}{c}{FWHM}
&\multicolumn{1}{c}{$\int T_{\rm mb}\delta \rm{v}$}\\
\multicolumn{1}{c}{} &\multicolumn{1}{c}{(km s$^{-1}$)}&\multicolumn{1}{c}{(km s$^{-1}$)}&\multicolumn{1}{c}{(K km s$^{-1}$)}\\
\hline
&\multicolumn{3}{c}{$^{13}$CO(6--5)}\\
hot core&$-44.74\pm0.03$&$7.23\pm0.06$&$107.8\pm0.8$\\
IRDC&$-45.99\pm0.02$&$5.44\pm0.05$&$67.3\pm0.8$\\
H{\sc ii}&$-48.97\pm0.01$&$8.40\pm0.02$&$202.2\pm0.5$\\
&\multicolumn{3}{c}{$^{13}$CO(8--7)}\\
hot core&$-44.63\pm0.06$&$6.5\pm0.2$&$43.5\pm0.9$\\
IRDC&$-46.5\pm0.2$&$4.4\pm0.6$&$16\pm2$\\
H{\sc ii}&$-48.83\pm0.05$&$8.2\pm0.1$&$116\pm1$\\
&\multicolumn{3}{c}{$^{13}$CO(10--9)}\\
hot core&$-44.28\pm0.07$&$6.4\pm0.2$&$24.4\pm0.5$\\
IRDC&$-46.8\pm0.1$&$3.7\pm0.3$&$4.7\pm0.3$\\
H{\sc ii}&$-49.56\pm0.04$&$7.67\pm0.09$&$52.4\pm0.5$\\
\hline
\end{tabular}
\end{table}

All data were smoothed to the resolution of the $^{13}$CO(10--9) map.
We selected three positions for the analysis: the hot core, the IRDC position  ((30\arcsec, 30\arcsec) from the centre of the APEX maps), 
and the centre of the H{\sc ii} region. 
 The IRDC position 
was selected to be a position associated with high column density in the infrared dark cloud (see Figs.~\ref{co}, \ref{13co} and \ref{colden}) 
but without IR emission. However, it is only $10\arcsec$ to the north of the EGO candidate \citep[][see Figs.~\ref{co}-\ref{13co}]{2008AJ....136.2391C}, and therefore, given the beam
of the observations, contamination from the embedded YSO may still be possible.  
Table~\ref{para} reports the measured line parameters of the $^{13}$CO transitions obtained with Gaussian fits; 
based on these values, we adopt line widths of 6, 3 and 7, for the hot core, the IRDC and the H{\sc ii} respectively, in agreement with values reported by \citet{sanjose} for the $^{13}$CO(10--9) line towards a sample of intermediate- and high-mass sources.
The spectra are shown in Fig.~\ref{spectra_13co}.

\begin{figure}
\centering
\includegraphics[angle=-90,width=9cm]{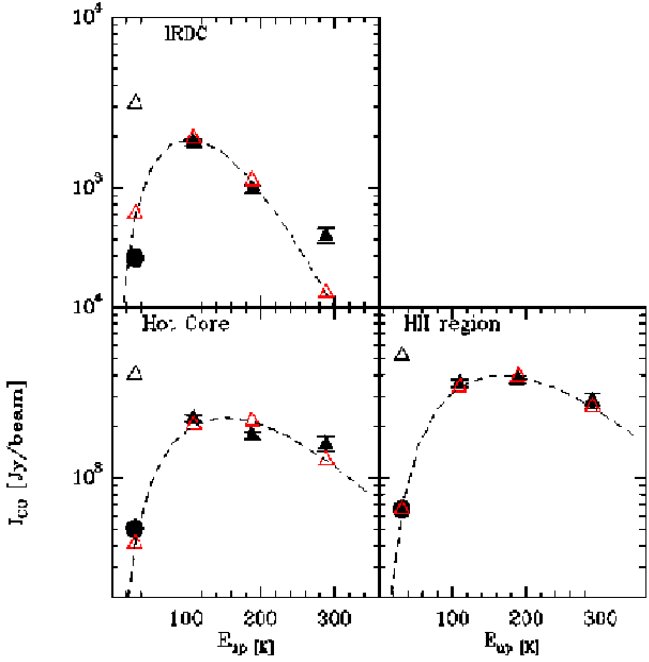}
\caption{Distribution of the $^{13}$CO peak line intensities. Full black triangles correspond to $^{13}$CO(6--5), (8--7) and (10--9) 
observed intensities. The circle represents the observed C$^{18}$O(3--2) flux,  the empty black triangle  the flux of the C$^{18}$O(3--2) line multiplied by 
$X_{\rm{^{13}CO}/\rm{C^{18}O}}\sim 8$. The red triangles are the best model fit results. The error bars 
include only calibration uncertainties. The dashed lines represent the best fit $^{13}$CO ladder.}
\label{sed_plot}
\end{figure}

The results of the RADEX analysis are listed in Table~\ref{fit} and shown in  Fig.~\ref{sed_plot}. For the IRDC position, the $^{13}$CO(10--9) line intensity is not well fitted by our one-temperature model. This likely reflects the fact that the $^{13}$CO(10--9) spectrum is dominated by the embedded source while the other two lines sample 
colder gas in the envelope. We considered a 20\% calibration error for the $^{13}$CO(6--5) and (8--7) observations and
a 15\% error for the $^{13}$CO(10--9) data.  \citet{2006A&A...454L..91W} mapped the region with the APEX telescope 
 in the C$^{18}$O(3--2) line. Therefore, since no observations were performed in the  $^{13}$CO(3--2) transition, 
we included the C$^{18}$O(3--2) data in Fig.~\ref{sed_plot}. Note however, that the C$^{18}$O(3--2) fluxes are not included in the fitting procedure, but that they are simply used to cross-check results.
Since the  C$^{18}$O(3--2) flux  corresponds to a lower limit to the flux of 
$^{13}$CO(3--2) line, we also plotted the  C$^{18}$O(3--2) flux corrected for the abundance ratio of $^{13}$CO to C$^{18}$O, 
$X_{\rm{^{13}CO}/\rm{C^{18}O}}\sim 8$ \citep{1994ARA&A..32..191W}. This value is likely an upper limit to the flux
of the  $^{13}$CO(3--2) line due to opacity effects.
  
\begin{figure}
\centering
\includegraphics[angle=-90,width=8cm]{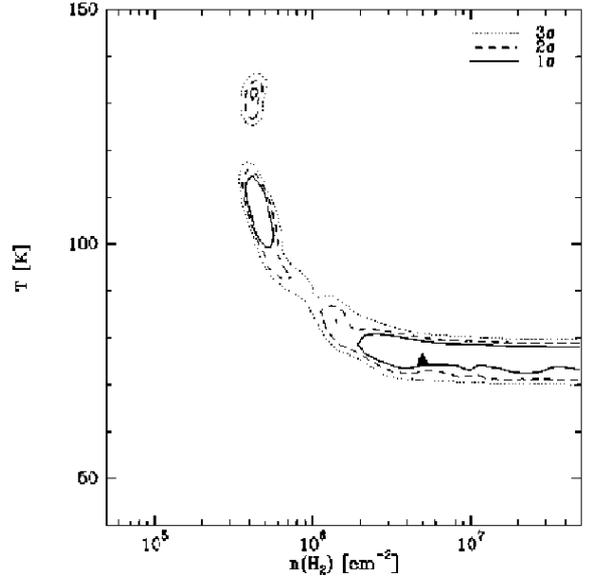}
\caption{Projection of the 3-dimensional (T-n-N) distribution of the $\chi^2$ 
on the $T-n$ plane for the H{\sc ii} position. The contours show the 1, 2 and 3$\sigma$ confidence 
levels for two degrees of freedom. The triangle marks the best fit position.}
\label{chi2_plot}
\end{figure}

 An example of the $\chi^2$ distribution projected on to the $T-n$ plane is shown in Fig.~\ref{chi2_plot}  for the H{\sc II} position, where the reduced $\chi^2$ at the best fit position is 3. Figure~\ref{chi2_plot} shows the typical inverse $n-T$ relationship often seen in $\chi^2$ distributions and due to the fact that density and temperature are, in the case of the CO molecule, not independent parameters \citep[see Appendix C of][]{2007A&A...468..627V}.

 We note here that the detection
of the $^{13}$CO(10--9) line breaks the degeneracy between density and temperature typical of $^{12}$CO analyses \citep[e.g.,][]{2004A&A...424..887K,2007A&A...468..627V} and help to give stronger constraints: indeed, with the exception of the hot core, at least the temperature of the gas is well determined at all positions.

 The results of the RADEX analysis 
 confirm that the LTE assumption used in Sect.~\ref{lte} is reasonable since the inferred densities are much larger
than the critical densities (a few $10^4$~cm$^{-3}$ for all three analysed transitions). Assuming 
$X_{\rm{^{12}CO}/\rm{^{13}CO}}\sim 60$ and an abundance of $^{12}$CO relative to H$_2$ of $2.7\times10^{-4}$, the derived
$^{13}$CO column densities listed in Table~\ref{fit} correspond to H$_2$ column densities of some $10^{22}$~cm$^{-2}$
for the hot core and the IRDC position, and of $2\times 10^{22}$~cm$^{-2}$
for the H{\sc ii} position.
The derived column densities and temperatures are in agreement with those derived in Sect.~\ref{lte} through rotational
diagrams of the $^{13}$CO emission.
The $^{13}$CO(6--5) optical depths are also in agreement with those estimated in Sect.~\ref{lte} through the $^{13}$CO and $^{12}$CO(6--5) line ratio, 
with the exception of the hot core
position: $\tau_{\rm{LTE}}\sim$1.2 and $\tau_{\rm{RADEX}}\sim$0.3 for the hot core, 
0.6 and 0.5 for the H{\sc ii} region, and 0.6 and 0.7 for the IRDC position.
\begin{table}
\begin{center}
\caption{Best fit parameters of the $^{13}$CO line modelling}\label{fit}
\begin{tabular}{cccc}
\hline
\hline
\multicolumn{1}{c}{Position}&\multicolumn{1}{c}{$T$}&\multicolumn{1}{c}{$n_{\rm{H}_2}$}&\multicolumn{1}{c}{$N_{\rm{^{13}CO}}$}\\
\multicolumn{1}{c}{} &\multicolumn{1}{c}{(K)}&\multicolumn{1}{c}{(cm$^{-3}$)}&\multicolumn{1}{c}{(cm$^{-2}$)}\\
\hline
hot core&$70(>60$)&$1\times 10^7 (>10^4)$&$5\times10^{16}$\\
IRDC&$45^{+20}_{-11}$&$1\times 10^6 (>10^5)$&$5\times10^{16}$\\
H{\sc ii}&$75^{+55}_{-7}$&$5\times 10^6 (>4\times10^4)$&$1\times10^{17}$\\
\hline
\end{tabular}
\end{center}
The errors represent the $3\sigma$ confidence levels in the temperature-density plane or  $3\sigma$ lower (shown in brackets) limit when no stronger constraints can be inferred.
\end{table}

 We finally stress that the results obtained with the 
LTE analysis (Sect.~\ref{lte}) and with the RADEX code (this section) are based on the assumptions that 1) all lines have a beam filling factor of one, and 2) that the emitting gas is homogeneous, whereas self-absorption profiles in the 
$^{12}$CO lines indicate an excitation gradient along the line of sights. For optically thin lines 
($^{13}$CO in the current case), a beam filling factor less than one (but equal for all transitions) 
would mostly affect the column density and result in larger values of $N$; for optically thick lines, 
a smaller value of $\eta$ would imply larger values of density and/or temperature. 

The uncertainties on the derived parameters due to the assumption of a homogeneous medium are 
of less immediate interpretation, and more complex radiative transfer codes 
\citep[e.g.,][]{2000A&A...362..697H} should be used to reproduce the observed line velocity profiles.

\section{Discussion}\label{dis}

\subsection{Total $^{12}$CO and $^{13}$CO emission}\label{tot}
\begin{figure*}
\centering
\includegraphics[angle=-90,width=18cm]{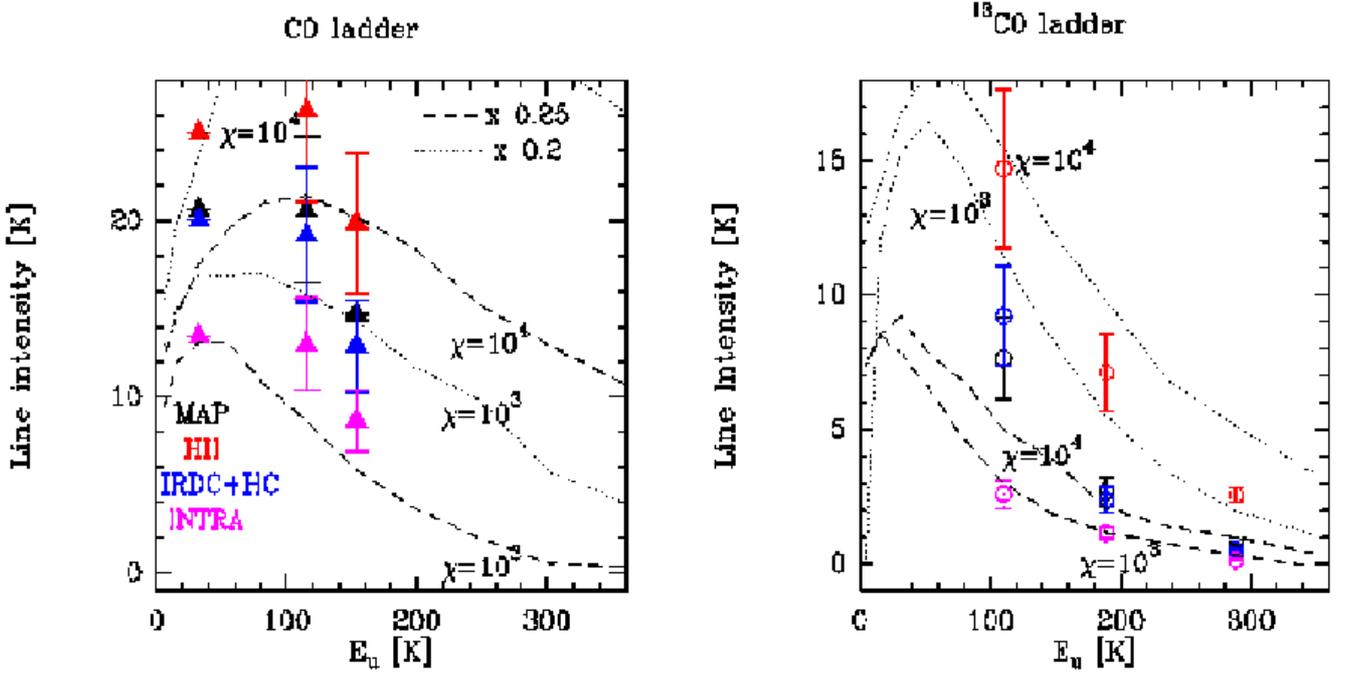}
\caption{
CO (left) and $^{13}$CO (right) ladders for the whole mapped region (black), the H{\sc ii} region G327.3--0.5 (red), 
the IRDC including the hot core (blue) and the inter-clump gas (magenta). Error bars include only calibration uncertainties.  In both panels, the dashed and dotted curves represent the predicted intensities for model B  
from \citet{1994A&A...284..545K} for a density of 10$^{7}$~cm$^{-3}$, incident UV fields of 10$^3$ and 10$^4$ relative to the average interstellar field}, a visual extinction of 10, and a Doppler broadening of 3 (dashed curve) and 1~km~s$^{-1}$ (dotted curve).
\label{ladders}
\end{figure*}

\begin{figure}
\centering
\includegraphics[angle=-90,width=8cm]{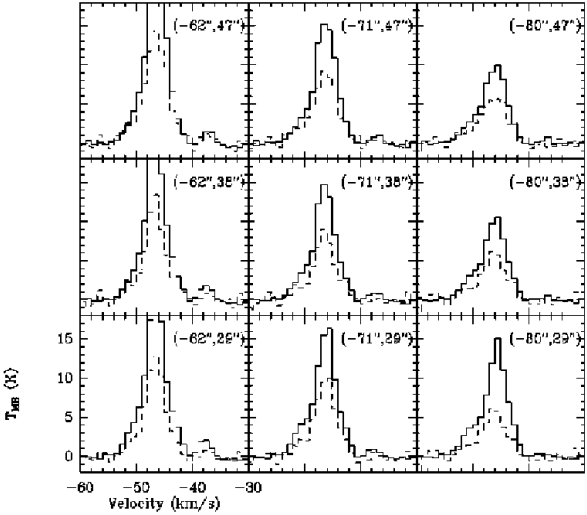}
\caption{CO(6--5) (solid line) and $^{12}$CO(7--6) (dashed line) spectra towards some positions in the inter-clump gas. The offset position from the centre of the APEX $^{12}$CO maps (Sect.~\ref{obs_a}) is shown for each spectrum in the top right corner.}\label{intra_co}
\end{figure}

Figure~\ref{ladders} shows the  $^{12}$CO and $^{13}$CO ladders obtained by averaging the emission of the different observed transitions, smoothed to the resolution of the HIFI data,
over four regions: the total map, the H{\sc ii} region 
G327.3--0.5, the IRDC hosting the hot core (the selected region does include the hot core), and finally the inter-clump gas, which was
defined as the region in the map where the $^{13}$CO lines are not 
detected on individual spectra. This region has  an equivalent radius of 50\arcsec (corresponding to $\sim$0.8~pc at the distance of the source).
Examples of $^{12}$CO line profiles towards the inter-clump gas are shown in Fig.~\ref{intra_co}. The C$^{18}$O(3--2) cannot be used in this analysis because the observations cover a much smaller
region than that mapped in $^{13}$CO. 
The total mass of the mapped region can be computed using  the excitation temperature and H$_2$ column density distributions 
shown in Figs.~\ref{tex} and \ref{colden}.  This corresponds to $\sim
700$~M$_\odot$.

The four regions have similar $^{12}$CO and $^{13}$CO ladders.  In order to correct for self-absorption,  
integrated fluxes  were obtained for each of the four regions
from line fitting of the CO spectra  with one Gaussian component. While this works fine for the spectra of the IRDC hosting the hot core
and for the inter-clump gas, the line profiles of the total map and of H{\sc ii} region are red-skewed and therefore the fluxes derived with this method are likely underestimated.
For the main isotopologue, the flux of the (7--6) and (6--5) 
transitions is very similar, although for the IRDC+HC and the inter-clump the flux of the (7--6) line is lower than that of the (6--5) line.
On the other hand, in the $^{13}$CO ladder the peak flux decreases with increasing energy level. 
For all transitions presented in this paper, the spectra are dominated in intensity  
by the H{\sc ii} region, 
whose flux is of the order of 
60\% of the total flux for the main isotopologue lines, and ranging
from $\sim 49\%$ of the total flux in the (6--5) line to $\sim 85\%$ in the (10--9) 
transition  for $^{13}$CO (see Table~\ref{percentage}).
The main difference between the $^{12}$CO spectra and those of $^{13}$CO lies in the emission from the inter-clump gas:
for all $^{12}$CO lines, the intensity is relatively strong, $10\%$ of the flux from the whole region. On the other hand,
 the flux of the $^{13}$CO lines coming from the inter-clump gas decreases with increasing J, from $\sim7\%$ of the total
flux for J=6 to $\sim 2\%$ for J=10.
The general behaviour of  the $^{12}$CO and $^{13}$CO ladders 
is qualitatively 
compatible with PDR models from 
\citet{1994A&A...284..545K} for high density ($10^6-10^7$~cm$^{-3}$)  
 clouds illuminated on one side by a UV radiation field (their model B). In Fig.~\ref{ladders}, we show 
the predicted CO and $^{13}$CO line intensities  for models with  a density of 10$^{7}$~cm$^{-3}$, 
incident UV fields with strength 10$^3$ and 10$^4$ relative to the average interstellar field \citep{1978ApJS...36..595D}, a visual extinction of 10, and a Doppler broadening of 3 and 1~km~s$^{-1}$. 
High densities ($n>10^6$~cm$^{-3}$, in agreement with our results from Sect.~\ref{sed}) 
are needed to locate the peak of the CO 
ladders at mid-, high-$J$ transitions (see Figs.~9--10, 12--13 of \citealt{1994A&A...284..545K}). Stronger UV radiation fields also shift the peak of the CO ladders to higher J transitions than  observed.
In  Fig.~\ref{ladders}, the $^{12}$CO model intensities are corrected by a factor 0.2 
and 0.25 to correct for different line-widths between the model and the observations, and possibly for 
beam dilution effects. On the other hand, 
 the $^{13}$CO results of Fig.~\ref{ladders} are not scaled down by any factor as they are  far too weak to match the observations.  
 \citet{1994A&A...284..545K}
 already noticed that the predicted 
line intensities of mid-$J$
$^{13}$CO transitions in their models are much weaker than observed in star-forming regions. They 
proposed that mid-$J$ $^{13}$CO emission comes from  
 a large number of filamentary structures, or clumps, 
along the line of sight. In this way, 
the modelled line intensity of mid-$J$ $^{13}$CO lines would increase significantly   as the lines  are optically thin, 
while it would not change for optically thick transitions.

Rotational diagrams of the $^{13}$CO emission
applied to the spectra of the H{\sc ii} region, of the IRDC and of the inter-clump 
gas infer rotational temperatures of 
66~K, 47~K and 44~K, respectively, and $^{13}$CO column densities of 
$6\times10^{15}$~cm$^{-2}$ for the H{\sc ii} 
region and the IRDC,  and  of $2\times10^{15}$~cm$^{-2}$ for the inter-clump gas. 
Since the inter-clump gas has physical parameters very similar 
to those of the IRDC region, but a much lower column density, 
we suggest that it is composed of high-density clumps with
low filling factors.   This is again in agreement with PDR models \citep[e.g.,][]{1994A&A...284..545K,2008A&A...488..623C} which predict 
strong emission at mid-$J$ $^{13}$CO and high-$J$ $^{12}$CO lines in the case of small, low mass, high density clumps.
 
\begin{table}
\centering
\caption{Percentage of integrated fluxes from the H{\sc ii} region G327.3--0.5 and from the inter-clump gas respect 
 to the total integrated flux in the map.}\label{percentage}
\begin{tabular}{rrr}
\hline
\hline
\multicolumn{1}{c}{Line} &\multicolumn{1}{c}{H{\sc ii}}&\multicolumn{1}{c}{Inter}\\
\hline
CO(3--2)&56\%&10\%\\
CO(6--5)&58\%&10\%\\
CO(7--6)&60\%&10\%\\
$^{13}$CO(6--5) &49\%&7\%\\
$^{13}$CO(8--7) &67\%&5\%\\
$^{13}$CO(10--9)&85\%&2\%\\
\hline

\end{tabular}
\end{table}

\citet{2008A&A...488..623C} suggested that the COBE  $^{12}$CO ladder of the Milky Way can be reproduced by a clumpy PDR model, 
and that the bulk of the Galactic FIR line emission comes from PDRs around the Galactic population of massive stars. Our
observations seem to confirm this result, since the CO emission of the G327.3--0.6 region is dominated by the PDR around
the H{\sc ii} region. Our results are also consistent with  the findings from  
\citet{2011MNRAS.tmp.1015D} and \citet{2011ApJ...730L..33M}.  These authors studied the properties
of massive  YSOs and compact H{\sc ii} regions in the RMS survey \citep{2005IAUS..227..370H}, and found   
that there is no significant population of massive YSOs above $\sim10^5~L_\odot$, while compact 
H{\sc ii} regions are detected up to $\sim10^6~L_\odot$. Since high-$J$ CO lines  are among the most important cooling lines in PDR, they reflect the luminosity of their heating sources: if the luminosity distribution of massive 
stars in the Galaxy is dominated by H{\sc ii} regions and not by younger massive stars,  
then we also expect that the CO distribution follows the same rule.

\subsection{Comparisons with other star forming regions}
Large-scale mapping of some low- and high-mass star forming regions
was performed in several $^{12}$CO transitions. However, given the different
critical densities of the lines, we prefer to compare our results with
studies carried on with J$>3$~CO lines. Excitation temperatures around
8--30~K are found in extended diffuse emission and towards the
brightest positions, respectively, in low-mass star forming regions
\citep[e.g.,][]{2010MNRAS.405..759D, 2010MNRAS.401..455C}, while
massive star-forming regions are usually warmer as confirmed by our
findings \citep[see
  also][]{2001ApJ...557..240W,2007A&A...461..999J,2010A&A...518L..82K,2011ApJ...728...61W,peng}. The
CO column densities ($10^{17}-10^{18}$~cm$^{-2}$ from the LTE analysis) found in G327 are
also comparable with values obtained in other massive star-forming
regions
\citep[][]{2001ApJ...557..240W,2007A&A...461..999J,2010A&A...518L.114W,2011ApJ...728...61W}.

Comparison with $^{12}$CO large maps of other high-mass star forming regions would be important to verify whether our result that
the $^{12}$CO distribution is dominated by the H{\sc ii} region is a common feature or not. However, most studies do not cover
different evolutionary phases as in our case. For OMC-1, \citet{peng}  confirmed
that the peak of the integrated intensity of several CO isotopologue lines is 
close to the Orion-KL hot core (although Orion-south and the Orion Bar PDR  are also
very prominent). One should notice however, that Orion is  roughly six times closer to the Sun than G327.3--0.6, and therefore the Orion Bar and Orion-KL would be much closer on sky ($\sim 30\arcsec$) if one would place them at the distance of
 G327.3--0.6.  
Moreover, Orion-KL likely represents a special case since it hosts a very powerful outflow \citep[e.g.,][]{1976ApJ...210L..39K,1984ApJ...284..176S}, which could alter the distribution of $^{12}$CO in the region. Indeed, \citet{2004ApJ...612..940M} show that broad velocity
emission arises mainly from the Orion-KL
region, while much of the narrower emission
arises from the PDR excited by the M42 H{\sc ii} region.

\subsection{Self-absorption profiles}
 As noted in Sect.~\ref{velo}, all $^{12}$CO transitions analysed in this paper are affected by self-absorption, 
which is likely to be due to cold gas surrounding a warmer component \citep{1981ApJ...245..512P}. 
 In particular along the infrared dark cloud (Fig.~\ref{hc}), the $^{12}$CO(6--5) line has blue-skewed profiles in 
the north-east (towards the EGO and the IRDC positions) and the red-skewed ones towards
the south-west (the hot core).
Blue- and red-skewed profiles  \citep[e.g.,][]{1997ApJ...489..719M} are commonly interpreted as due to rotation or outflow motions (which 
should produce equal numbers of red and blue profiles, and could therefore explain the profiles detected towards 
the EGO and IRDC position, the hot core and the H{\sc ii} region, Fig.~\ref{spectra}) or to infall (which 
should produce profiles which are skewed towards the blue, e.g. towards the EGO and IRDC position) or to expansion (which should produce  
profiles skewed
towards the red and could be responsible for the $^{12}$CO spectra of the hot core and the H{\sc ii} region). 
From the PV diagram of the $^{12}$CO(6--5) line, we do not have any evidence of global rotation towards 
the hot core and the red-skewed profile can be interpreted in terms of expansion or outflow motion. Similarly, 
from Fig.~\ref{shell} we see that the emission does not  follow a perfect expanding spherical shell 
which might imply that the rotation is on the origin of the self-absorbed profile seen towards the H{\sc ii} region. Finally, the blue-skewed line profile detected toward the EGO and IRDC positions are typical of infall motion.

\section{Conclusions}\label{end}

To study the effect of feedback from massive star forming regions in their surrounding environment, we selected the region
G327.3--0.6 for large scale mapping of several mid-$J$ $^{12}$CO and $^{13}$CO lines with the APEX telescope and of the high-$J$ 
$^{13}$CO(10--9) transition with the {\it Herschel} satellite. Our results can be summarised as follows:
\begin{enumerate}
\item Maps of all transitions are dominated by the PDR associated with the H{\sc ii} region G327.3--0.5; mid-$J$ $^{12}$CO and $^{13}$CO emission is detected along the whole extent of the IRDC;
\item Mid-$J$ transitions show  rather extended emission with typical excitation temperatures of $\sim 30$~K and column densities of some $10^{21}$~cm$^{-2}$;
\item All observed transitions are detected also in the inter-clump gas when averaged over large regions. 
The inter-clump  emission is compatible with LTE emission from a 
gas at 44~K, and with  a $^{13}$CO column density of 2$\times$10$^{15}$~cm$^{-2}$, thus suggesting that the inter-clump  is composed of high-density clumps with low filling factors;
\item The warm gas traced by $^{12}$ and $^{13}$CO(6-5)
is only a small percentage ($\sim$10\%) of the total gas in the infrared dark cloud, while it reaches values up to 
$\sim$35\% of the total gas in the ring surrounding the H{\sc ii} region;
\item The $^{12}$CO and $^{13}$CO ladders
are qualitatively compatible with PDR models for high density gas ($n>10^6$~cm$^{-3}$) and, in the case of $^{13}$CO, suggest
that the emission comes from  
 a large number of clumps;
\item The detection of the $^{13}$CO(10--9) line allows to give stronger 
constraints on the physics of the gas by breaking the degeneracy between density and temperature (typical of $^{12}$CO and $^{13}$CO transitions) in the high temperature--low density part of the $T-n$ plane. 
\end{enumerate}

\begin{acknowledgements}
The authors thank Dr. Joe Mottram for a careful review of the manuscript and an anonymous referee for useful comments and suggestions.

{\it Herschel} is an ESA space observatory with science instruments provided
by European-led Principal Investigator consortia and with important
participation from NASA. 
HIFI has been designed and built by a consortium of
 institutes and university departments from across Europe, Canada and the
 United States under the leadership of SRON Netherlands Institute for Space
 Research, Groningen, The Netherlands and with major contributions from
 Germany, France and the US. Consortium members are: Canada: CSA,
 U.Waterloo; France: CESR, LAB, LERMA, IRAM; Germany: KOSMA,
 MPIfR, MPS; Ireland, NUI Maynooth; Italy: ASI, IFSI-INAF, Osservatorio
 Astrofisico di Arcetri- INAF; Netherlands: SRON, TUD; Poland: CAMK, CBK;
 Spain: Observatorio Astron{\'o}mico Nacional (IGN), Centro de
 Astrobiolog{\'i}a
 (CSIC-INTA). Sweden: Chalmers University of Technology - MC2, RSS $\&$
 GARD; Onsala Space Observatory; Swedish National Space Board, Stockholm
 University - Stockholm Observatory; Switzerland: ETH Zurich, FHNW; USA:
 Caltech, JPL, NHSC.

\end{acknowledgements}

\end{document}